\documentclass[12pt]{article} 
\usepackage{amsmath,amsthm,amsfonts,amssymb}

\def\be{\begin{equation}}\def\ee{\end{equation}}
\def\bea{\begin{eqnarray}}\def\eea{\end{eqnarray}}
\def\p{\partial}
\def\IP{\mathbb{P}}

\def\eps{\epsilon}
\def\DL{\mathfrak{L}}\def\qq{\mathfrak{q}}
\def\td{\textrm{td}}
\def\qcor#1{\langle #1 \rangle}\def\qcord#1{\langle\!\langle #1 \rangle\!\rangle}
\def\cx#1{{\cal#1}}
\def\Iv{\mathcal{I}}\def\ngv{\mathfrak{n}}\def\vd{{\vec{d}}}\def\that{\hat t}
\mathchardef\mm="2D

\begin{document}
\begin{titlepage}
\begin{flushright}
{\tt BONN-TH-2019-03}\\
{\tt LMU-ASC 19/19}\\[14ex] 
\end{flushright}
\begin{center}
{\Large\bf Quantum K-Theory of Calabi--Yau Manifolds}\\[6ex]
{\Large Hans Jockers$^{1}$ and Peter Mayr$^{2}$}\\[6ex]
$^1\,$Bethe Center for Theoretical Physics,\\
Physikalisches Institut, Universit\"at Bonn, 53115 Bonn, Germany\\
{\tt jockers@uni-bonn.de} \\[2ex]
$^2\,$Arnold Sommerfeld Center for Theoretical Physics,\\
Ludwig-Maximilians-Universit\"at,
80333 Munich, Germany\\
{\tt mayr@physik.uni-muenchen.de} \\[6ex]
\begin{abstract} The disk partition function of certain 3d $N=2$ supersymmetric gauge theories computes a quantum K-theoretic ring for K\"ahler manifolds $X$. We study the 3d gauge theory/quantum K-theory correspondence for global and local Calabi--Yau manifolds with several K\"ahler moduli. We propose a multi-cover formula that relates the 3d BPS world--volume degeneracies computed by quantum K-theory to Gopakumar--Vafa invariants.
\end{abstract}
\end{center}
{\tt
\noindent
}
\vfill
\noindent
May 2019
\end{titlepage}

\section{Introduction and Outline}\def\markt{$\spadesuit$}
The works of Nekrasov \cite{Nekphd} and  Nekrasov and Shatashvili \cite{NekSha2,NekSha} establish, amongst many others,  a fundamental relation between 3d supersymmetric gauge theories compactified on a circle and quantum K-theory on the moduli space. For the concrete case of massless theories with a non-trivial UV-IR flow, a 3d gauge theory/quantum K-theory correspondence connecting the BPS partition functions of certain $\cx N=2$ supersymmetric gauge theories to Givental's permutation equivariant quantum K-theory \cite{Giv15all} on K\"ahler manifold $X$ was proposed and studied in ref.~\cite{JM}. It lifts the correspondence between the $\cx N=(2,2)$ 2d gauged linear sigma model (GLSM) and quantum cohomology \cite{WitPhases,MP} to 3d world-volumes of topology $\Sigma\times S^1$, in line with the results of refs.~\cite{Nekphd,NekSha2}. The cohomological quantum product is lifted to a K-theoretic quantum product of bundles on the moduli space of stable maps from $\Sigma$ to $X$, related to the action of Wilson line operators \cite{Dimofte:2011py,KW13,BDP,CV13} in the $\cx N=2$ 3d gauge theory.\footnote{For a discussion of quantum K-theory in the context of $\cx N=4$ supersymmetric theories see refs.~\cite{NekSha2,NekSha,GK,Bullimore:2016nji,Aganagic:2016jmx,Aganagic:2017gsx,KPSZ,Bullimore:2018jlp}.}  In the other direction, the 2d quantum cohomology can be recovered from the small radius limit of the 3d theory. It appears that the 3d correspondence is more fundamental  than its better known 2d limit in several aspects. For example,  integrality of the coefficients of the instanton expansions is manifest in 3d, due to their interpretation as BPS degeneracies on the world-volume, or as holomorphic Euler numbers in the quantum K-theory of ref.~\cite{Giv15all}. This interpretation applies also to the coefficients of the mirror map \cite{JM}, giving a physical derivation of the integrality properties proven before in ref.~\cite{Lian} for the quintic by different methods.

In this note we continue to study the 3d correspondence in the special case where $X$ is a Calabi--Yau manifold,  which is the natural setup for string and M-theory. For Calabi--Yau threefolds we describe a closed formula that relates the degeneracies of 3d world--volume operators to the degeneracies of BPS states in the 5d target space, as counted by the M-theoretic genus zero Gopakumar--Vafa invariants \cite{GV98}. The difference between the counting of BPS objects on the 3d world-volume and of 5d BPS objects in target space shows up in the contribution of multi--covers in $n$-pt functions with $n<3$. The presented 3d world-volume BPS counting arguments apply for higher-dimensional Calabi--Yau manifolds (and higher genera) as well, even if a M-theoretic target space interpretation is not available. From this perspective, the 3d world-volume BPS degeneracies are more universal. It would be interesting to uncover the physics origin of the relationship between the 3d world-volume and the target space BPS indices.

\section{Multi-Cover Formula for Calabi--Yau Threefolds} \label{sec:MCSec}

\subsection{Complete Toric Intersections}
In the following we outline the computation of quantum K-theoretic invariants for $\cx N=2 $ 3d gauge theories with Higgs branches corresponding to complete intersection Calabi--Yaus (CICY) in toric hypersurfaces with several K\"ahler moduli. As a concrete example we consider the Calabi--Yau threefold $X=\IP^4_{1,1,2,2,2}[8]$, defined as the proper transform of the zero locus of a degree eight polynomial in a smooth resolution of the weighted projective space $\IP^4_{1,1,2,2,2}$. This is a Calabi--Yau  hypersurface with $n_K=2$ K\"ahler moduli that has been studied  in much detail in the context of 2d mirror symmetry  in refs.~\cite{Cand,HKTY}.

\subsubsection*{Difference Equations}
As in ref.~\cite{JM}, we consider the $\cx N=2$ supersymmetric 3d lift of the GLSM with gauge group $U(1)^{n_K}$ with $n$ charged matter fields with charges $q_{a\alpha}$, representing the homogeneous coordinates of the toric ambient space $W$. In addition there are $\ell$ fields of negative charge $-d_{a\beta}$ corresponding to the hypersurface constraints of degrees $d_{a\beta}$. These data are collected in the charge vectors $l_a = (-d_{a\beta};q_{a\alpha})$, $a=1,\hdots n_K$, $\alpha \in N$, $\beta\in D$. The index sets $N$ and $D$  refer to the fields with Neumann and Dirichlet boundary conditions in the 3d GLSM on $S^1 \times_q D^2$. The Ward identities satisfied by the 3d Wilson line operators associated with the Calabi--Yau manifold $X$ can be represented by the difference operators
\bea\label{DiffEqgen}
\DL_a &=&  
\prod_{\substack{\alpha\in N\\l_{a\alpha>0}}}\prod_{j=0}^{l_{a\alpha}-1}(1-q^{\vartheta_\alpha-j})\\&&-Q_aq^{\frac  12 A_{aa}+B_a} 
q^{\sum_i A_{ai}\theta_i}\prod_{\beta\in D}\prod_{j=1}^{|l_{a\beta}|}(1-q^{-\vartheta_\alpha+j}) \,
\prod_{\substack{\alpha\in N\\l_{a\alpha}<0}}\prod_{j=0}^{|l_{a\alpha}|-1}(1-q^{\vartheta_\alpha-j})\ .\nonumber
\eea
Here $A_{ab}$ and $B_a$ parameterize the (effective) Chern--Simons levels in the 3d gauge theory and 
\be\textstyle
\vartheta_\alpha = \sum_a l_{a\alpha} \theta_a\ ,\qquad \theta_a = Q_a\frac{\p}{\p Q_a}\ ,\nonumber
\ee
where $Q_a$ are the exponentials of the Fayet-Iliopoulos parameters.
The difference operators $\DL_a$ annihilate the  $S^1 \times_q D^2$ partition function for supersymmetric choice of boundary conditions \cite{JM}. The partition function takes the form of a multi-residue integral over Wilson line variables $\eps_a$ after supersymmetric localization \cite{BDP,Gadde:2013wq,YS}. The $Q$-dependent part of the integrand represents a vortex sum $\Iv$ satisfying $\DL_a \Iv = 0$, $a=1,\hdots n_K$. For the case of a 3d gauge theory with Higgs branch a K\"ahler manifold $X$, the expansion of $\Iv$ around the large volume limit $Q_a=0$ is a generalized $q$-hypergeometric series \bea
\Iv&=&c_0 \sum_{0\leq d_a \in \mathbb{Z}^{n_K}} Q^{\tilde d} \ c(\tilde d)\ q^{CS(d,\eps)}\, ,\nonumber
\eea
where $ \tilde d_a = d_a-\eps_a$, $Q^d= \prod_a Q_a^{d_a}$ and 
\bea\textstyle
c(d)&=&\frac{\prod_{\beta \in D}\Gamma_q(1-l_{a\beta}d_a)}{\prod_{\alpha\in N}\Gamma_q(1+l_{a\alpha}d_a)}\, ,\nonumber
\eea
in terms of the $q$-Gamma function $\Gamma_q(x)$. Moreover, $c_0 = c(-\eps)^{-1} $ and 
\be\textstyle
CS(d,\eps)=\frac 12 A_{ab}((d_a-\eps_a)(d_b-\eps_b)-\eps_a\eps_b)+B_ad_a\ ,\nonumber
\ee
is a contribution from the Chern--Simons term in the 3d bulk theory \cite{JM}. \\

\noindent\underline{Example:}
For the example with the Higgs branch corresponding to the Calabi--Yau threefold $\IP^4_{1,1,2,2,2}[8]$, we consider the 3d GLSM with the gauge group $U(1)^2$ and charged chiral matter fields as summarized by the charge vectors 
$$
l_1= (-4;1,1,1,1,0,0),\qquad l_2 = (0;0,0,0,-2,1,1)\ .
$$
Then the difference operators are
\bea\textstyle
\DL_1 &=& (1-p_1)^3(1-p_1p_2^{-2})-Q_1q^{\frac 12 A_{11}+B_1}\prod_a p_a^{A_{1a}} \prod_{i=1}^4(1-p_1^4q^i)\ ,\nonumber\\[-4mm]
\DL_2 &=& (1-p_2)^2-Q_2 q^{\frac 12 A_{22}+B_2}\prod_a p_a^{A_{2a}} \prod_{i=0}^1(1-p_1p_2^{-2}q^{-i})\ ,\nonumber
\eea
with $p_a=q^{\theta_a}$. The first operator can be factorized as $\DL_1= (1-p_1)\DL'_1$ with 
$$
\DL'_1 =   (1-p_1)^2(1-p_1p_2^{-2})-Q_1q^{\frac 12 A_{11}+B_1}\prod_j p_j^{A_{1j}} (\sum_{i=0}^3(qp_1)^i)\prod_{i=1}^3(1-p_1^4q^i)\ .\nonumber
$$
$\DL'_1$  will be the operator that annihilates only the $q$-periods on the hypersurface $X$, as opposed to that on the toric ambient space $W$. 

 \subsubsection*{Basis of Solutions}
Analogous to the Froebenius method for differential equations, the Taylor expansion of $\Iv$ in the Wilson line parameters $\eps_a$ generates a set of $N=$~dim$(K(X))=2(1+n_K)$ linearly independent solutions to the difference operators $\DL_a$, where $K(X)$ is the K-group of $X$.\footnote{By loosely referring to $K(X)$ we really mean the torsion-free K-theory group $K^0(X)\otimes \mathbb{Q}$ with coefficients in $\mathbb{Q}$. Using the Chern isomorphism we can then identity the torsion-free K-theory classes with cohomology classes in $H^{2*}(X,\mathbb{Q})$. Therefore, in the following, the notation $\Phi_A$ will denote both the K-theory element and its Chern character, depending on the context.}  Here linear independence is defined with respect to coefficients invariant under the $n_K$ shifts $p_aQ_b = Q_bq^{\delta_{ab}}$. In the 2d case, a geometric way to construct a vector of solutions is to study the central charges of D-branes. The  generalization to E-branes in the 3d GLSM proposed in ref.~\cite{JM} starts from the geometric interpretation of the residue integral as an integral over $X$ after the replacement 
\be\label{epsrep}
q^{-\eps_a}=P_a= e^{-\beta K_a}\ ,\qquad q=e^{-\hbar \beta}\ ,
\ee
where $K_a\in H^2(X,\mathbb{Z})\simeq H_4(X,\mathbb{Z})$ are the generators of the K\"ahler cone dual to the Mori cone defined by the charge vectors $l_a$.\footnote{We refer for instance to the book \cite{CK} for background material.} The parameter 
$\beta$ is the size of the $S^1$ and $\hbar$ is the weight of the twist of the geometry $S^1 \times_q D^2$ \cite{BDP}. The 2d limit is defined as $\beta\to 0$. Except for the unusual normalization by  an extra factor $\beta$, which comes from the extra circle in the 3d theory, $P_a$ is the (Chern character of the) line bundle $\cx O(-K_a)$ on $X$.

The $q$-series $\Iv$ with values in $K(X)$ obtained from the vortex sum by the replacement \eqref{epsrep} agrees,  for $A_{ab}=0=B_a$ and up to an overall factor,  with the $I$-function for the permutation symmetric quantum K-theory  defined in ref.~\cite{Giv15all}. The  case with non-zero $A_{ab},B_a$ is also interesting and includes the more general setup of quantum K-theory at higher level studied in ref.~\cite{RZ18}.  

To obtain a basis of $N$ linearly independent solutions we start with the ring relations among the cohomology elements $K_a$ on the toric intersection $X$. These are obtained from the construction of the toric ambient space $W$ as a GIT quotient $W=(\mathbb{C}^{\textrm{dim} W+n_K}-\Delta)/\!/(\mathbb{C}^*)^{n_K}$ in a standard way. We refer again to ref.~\cite{CK} for details. The result is the cohomology ring  
\be\label{DeltaK}\textstyle
H^{2*}(W,\mathbb{Z}) = \mathbb{Z}[K_a]/\Delta_K\ ,
\ee
where $\Delta_K$ is the ideal of relations. For simplicity we assume that the cohomology ring $H^{2*}(X,\mathbb{Q})$ can be generated by products 
\be\textstyle
K_A=K_{\vec a} =\prod_{i=1}^k K_{a_i}\in H^{2*}(W,\mathbb{Z})\ ,\nonumber
\ee
of the generators $K_a$ of $H^{2*}(W,\mathbb{Z})$ restricted to the CICY $X$, i.e. that there are no non-toric classes.\footnote{\label{fn3} The restricted classes are in general multiples of the generators of $H^{2*}(X,\mathbb{Z})$ on the CICY $X$; the extra factors become immaterial when passing to coefficients in $\mathbb{Q}$.} Here $A$ runs over the appropriate set of vectors $\vec a$ with 
$\dim(\vec a)\leq\dim(X)$ specifing the monomials $K_A$ up to degree $\dim(X)$. The restriction of the ideal $\Delta_K$ to $X$ gives an ideal $\Delta_{X,K}$ that reduces the set of $\sum_{i=0}^{\textrm{dim } X} {n_K \choose i}$ monomials $K_{\vec a}$ to a basis $\{K_A\}$ of dimension $N=\dim(K(X))$. 

To this cohomolgical basis we assign elements $\Phi_{\vec a}\in K(X)$ with Chern character\footnote{The normalization factor $\beta$ from the radius of the extra $S^1$ is kept to make the 2d limit manifest. To stay within $\mathbb{Q}$-cohomology, $\beta$ should be restricted to rational values.}
\be\label{Pdef}\textstyle
\Phi_a = 1-e^{-\beta K_a}=\beta K_a + O(\beta^2)\ , \qquad \Phi_{\vec a}=\Phi_{a_1\hdots a_k} =\prod_{i=1}^k \Phi_{a_i}\ .
\ee
One has $\Phi_{a_1\hdots a_k}\in H^{\geq 2k}(X,\mathbb{Q})$. Replacing $K_a\to \Phi_a$ in $\{K_A\}$ gives a basis $\{\Phi_A\}$ for $K(X)$. Expanding $\Iv$ in $\eps_a$ up to order $\dim(X)$ and reexpressing the result in terms  of the basis $\{\Phi_A\}$ gives
\be\textstyle
\Iv = 
\Iv_0\cdot 1 +\sum_{a=1}^{n_K} \Iv_a \Phi_a +\sum_{a\leq b} \Iv_{ab} \Phi_{ab} +\sum_{a\leq b \leq c} \Iv_{abc} \Phi_{abc} = 
\sum_{A=0 }^{N-1} \Iv_A\Phi_A\ .\nonumber
\ee
The coefficients $\Iv_A$ provide a basis of $N$ solutions to the difference equations.

In the 2d limit $\beta\to 0$, the difference operators \eqref{DiffEqgen} reduce to the GKZ differential operators of 2d mirror symmetry, and the $N$ solutions to the difference equations reduce to the ordinary periods associated with the $\cx N=2$ special geometry of the moduli space. A natural vector of solutions to the difference equations, which reduces to the standard solution vector used in 2d mirror symmetry \cite{HKTY2} in the small radius limit,\footnote{More precisely, the $A$-model is obtained for the special value $\hbar=-2\pi i$.} is
\be\label{Piq}
  \Pi_q(Q,q) = 
  \left(\begin{smallmatrix} 
    \Iv\\ -\frac 1 \hbar \p_{K_a}\Iv\\\frac 1 {2\hbar^2} \kappa_{abc}\p_{K_b}\p_{K_c}\, \Iv\\
    -\frac 1 {3!\hbar^3} \kappa_{abc}\p_{K_a}\p_{K_b}\p_{K_c} \, \Iv
   \end{smallmatrix}\right)
   = \left(\begin{smallmatrix} 
   1\\ -\frac 1 \hbar \ln Q_a\\ \frac 1 {2\hbar^{2}} \, \kappa_{abc}\ln Q_b\ln Q_c \\
   -\frac 1 {3!\hbar^{3}} \kappa_{abc}\ln Q_a \ln Q_b\ln Q_c
   \end{smallmatrix}\right) \cdot (1+O (Q)) \ .
\ee
Here $\kappa_{abc}=\int_X K_aK_bK_c$ are the intersection numbers and we used the relations
$
q=e^{-\beta \hbar}$, $\eps_a = -K_a/\hbar$.
The solution vectors in the basis $\{K_A\}$ and the basis $\{\Phi_A\}$ are related by the linear transformation $\Iv=\sum_A\Pi_{q,A}K_A = \sum_A \Iv_A\Phi_A$. The indices on the basis elements $\Phi_A$ are raised and lowered with the standard inner product 
\be\textstyle
(\Phi_A,\Phi_B)_K= \beta^{-\textrm{dim } X}\int_X \Phi_A \Phi_B\,  \td X =:\chi_{AB}\in\mathbb{Z}\ . \ee

\noindent \underline{Example:}
In the simple example, the ideal $\Delta_K= \{K^3_1(K_1-2K_2),K_2^2\}$ also follows from restricting the difference operators $\DL_a$ to the degree zero terms of $\Iv$. On the hypersurface $X\subset W$ one may drop one of the $K_1$ factors in the first entry of $\Delta_K$. A set of the $N=6$ basis elements for $H^{2*}(X,\mathbb{Z})$ is then
$$
\{K_A\} := \{\, 1;\, K_1,\, K_2;\, K_1^2,\, K_1K_2 ;\, K_1^2K_2\, \}\ , \quad A=0,\hdots 5\ .
$$
The intersections are $\kappa_{111}=8$, $\kappa_{112}=4$, $\kappa_{122}=\kappa_{222}=0$. The inner product on  the basis $\{\Phi_A\}$ induced by the replacement $K_a\to \Phi_a$ is\footnote{The overall factor $\tfrac1{16}$ in $\chi^{-1}$ is related to the comment in fn.~\ref{fn3}.} 
\be
\begin{tiny}
\chi_{AB} = \left(
\begin{array}{cccccc}
 0 & 6 & 2 & -8 & -2 & 4 \\
 6 & -8 & -2 & 8 & 4 & 0 \\
 2 & -2 & 0 & 4 & 0 & 0 \\
 -8 & 8 & 4 & 0 & 0 & 0 \\
 -2 & 4 & 0 & 0 & 0 & 0 \\
 4 & 0 & 0 & 0 & 0 & 0 \\
\end{array}
\right) \ ,
\quad 
\chi^{AB}=\frac 1 {16} \left(
\begin{array}{cccccc}
 0 & 0 & 0 & 0 & 0 & 4 \\
 0 & 0 & 0 & 0 & 4 & 2 \\
 0 & 0 & 0 & 4 & -8 & 4 \\
 0 & 0 & 4 & 0 & 2 & -1 \\
 0 & 4 & -8 & 2 & 0 & 2 \\
 4 & 2 & 4 & -1 & 2 & -6 \\
\end{array}
\right)
\end{tiny} \ .\nonumber
\ee

\subsubsection*{K-Theoretic Mirror Map}
The 3d partition function of the supersymmetric gauge theories is written in the UV variables. To connect this expression to enumerative invariants of ordinary quantum K-theory one needs to determine the flat coordinates in the IR theory. This step is often called the mirror map in the context of mirror symmetry. The K-theoretic version of the mirror map has been described by Givental in refs.~\cite{Giv15all, GivER} as a motion on the Langrangian cone in the symplectic loop space $\cx K$ described below. We continue with an  outline of the computation in two steps. The first step corresponds to removing the multi-trace deformations in the UV theory, and to shifting the input of the symmetric quantum K-theory to zero. The second step corresponds to a deformation by single trace operators in the gauge theory and determining the flat coordinates for these directions.

As argued in \cite{JM}, the vortex sum $\Iv(Q,q)$ of the 3d gauge theory takes value in the symmetric quantum K-theory of ref.~\cite{Giv15all}. More precisely it is related to Giventals $I$-function\footnote{One distinguishes the concepts of the Givental $I$-function and the $J$-function, which can be defined using localization methods in the moduli space of quasi-maps and stable maps, respectively. These  correspond to the GLSM and the non-linear sigma model, respectively. The two are related by a UV/IR reparameterization, which is part of the 3d mirror map discussed below. Despite the fact, that the correlator notation introduced in the following equations refers to the stable map compactification, i.e., the IR phase, we will continue to use the symbol $I$ for the generating function throughout this note also for the IR data.} by the relation
$$
\textstyle I_{SQK}(Q,q)=(1-q)\prod_a P_a^{-\ln Q_a/\ln q}\Iv(Q,q)\ .
$$ 
In the symmetric quantum K-theory, the deformations $t_A$ are characterized by the expansion
\be\textstyle
I_{SQK}(t)=(1-q)+t_A(Q,q)\Phi_A +\sum_{\vd>0}Q^\vd \qcord{\frac{\Phi^A}{1-qL}}_\vd\, \Phi_A\ .\nonumber
\ee
Here $\qcord{...}=\sum_{n\geq 0 } \frac 1 {n!} \qcor{...,t^n}$ denotes the correlator part with deformation $t=t_A\Phi_A$. The split into the input $t$ and the correlator part is defined by the decomposition $\cx K
=\cx K_+\oplus \cx K_-$ where\footnote{$\cx K$ is the symplectic loop space with 
pairing $\Omega(f,g) = (\operatorname{Res}_{q=0} + \operatorname{Res}_{q=\infty}) \frac{dq}{q} \left(f(q),g(q^{-1})\right)_K$. $R(q)$ denotes the field of rational functions in the variable $q$. $\mathcal{K}_\pm$ are Lagrangian subspaces of $\mathcal{K}$ with respect to the symplectic pairing, see refs.~\cite{GivTon, GivER}.} 
$$
\begin{aligned}
\ \ \mathcal{K} \,&=\,  K(X) \otimes \mathbb{C}(q,q^{-1})\otimes\mathbb{C}[[Q]] \ , \qquad 
  \mathcal{K}_+\,=\,K(X) \otimes \mathbb{C}[q,q^{-1}]\otimes\mathbb{C}[[Q]] \ , \\
  \mathcal{K}_- \,&=\,K(X) \otimes 
     \left\{\,r(q) \in R(q) \,\middle|\, \text{$r(0)\ne\infty$ and $r(\infty)=0$} \right\} \otimes\mathbb{C}[[Q]] \ ,
\end{aligned}   
$$
such that $t\in \cx K_+$.
The input $t(Q,q)$ corresponds to a complicated deformation of the 3d theory with multi-traces of Wilson line operators \cite{JM}. To obtain correlators in the ordinary quantum K-theory, we first shift this input to zero, by applying Givental's transformation
\be\label{gtfs}
I(0)=\exp\left(\sum_{r>0}\frac{\Psi_r(\eps)}{r(1-q^r)}\right) \, I(t) = (1-q)+0+\sum_ A F^A(0)\Phi_A\ ,
\ee
where $F^A(0)$ are the correlator parts at zero deformation.
The input 
$\eps = \sum_A\eps_A(Q,q)\Phi_A(Pq^\theta)$ is determined as a series in $Q$ such that the r.h.s. holds. Here $\eps_a\in \cx K_+$ and $\Phi_A(Pq^\theta)$ is the operator obtained by replacing $P_a$ by $P_aq^{\theta_a}$ in $\Phi_A(P)$. The Adams operator $\Psi_r$ acts as 
$$\textstyle
\Psi_r(Q)=Q^r,\qquad \Psi_r(q)=q^r,\qquad \Psi_r(P_a)=P_a^r,\qquad \Psi_r(q^{\theta_a})=q^{r\theta_a}\ .\nonumber
$$
It is shown in refs.~\cite{Giv15all} that transformations of the form \eqref{gtfs} generate the deformations on the family of symmetric quantum K-theory parameterized by $t_A$ and $Q_a$.  In the 3d gauge theory this transformation arises upon integrating in massive charged degrees of freedom \cite{JM}. The series expansion is tedious in practice, but suited to a computation by a symbolic computer program to given order in $Q_a$.

In the second step, the $I$-function $I(0)$ can now be deformed again to obtain correlators with deformations of the ordinary, symmetric, or equivariant quantum K-theory \cite{Giv15all}. The deformation family of ordinary quantum K-theory corresponds to the deformation by single trace operators in the 3d field theory \cite{JM}. The $I$-function for the ordinary quantum K-theory with input $t_A$ is obtained by a transformation of the same type as \eqref{gtfs} \cite{GivER}, but restricted to the $r=1$ term in the sum:
\be
I_{QK}(t)=\exp\left(\frac{\eps(t)}{1-q}\right) \, I(0) =(1-q)+ \sum_A t_A\Phi_A+\sum_{A} F_{QK}^A(t) \Phi_{A}\ .\nonumber
\ee
In the following we will restrict to $q$-independent deformations $t_A\in \mathbb{C}$, which corresponds to a deformation by operators with (effective) spin zero in the 3d gauge theory, or to setting the deformations in the direction of gravitational descendants to zero in the quantum K-theory.
\\

\noindent\underline{Example:} Performing the two steps in the example $\IP^4_{1,1,2,2,2}[8]$ we find the general form (dropping the subscript $QK$ again)
\be\label{I3f}\textstyle
I(t_A) = (1-q)+t_A\Phi_A+\sum_{A>n_K}F^A(t)\Phi_A\ .
\ee
The correlators $F^A$ are zero in the first 1+$n_K$ directions $\Phi_0=1$ and $\Phi_a=(1-P_a)$, as in the cohomological case. Moreover, the correlators depend, except for a classical, $Q$-independent term, only on the deformations $t_0\mathbf{1} +\sum_{a=1}^{n_K}t_a\Phi_a$. Since the dependence on $t_0$ is universal and determined by the K-theoretic string equation \cite{Lee:2001mb}, the enumerative invariants will be encoded in the quantum correlators $F^A(t_a)$ as functions of the deformations $\sum_{a=1}^{n_K}t_a\Phi_a$.

It is useful to express the correlators in terms of the dual basis 
\be\label{fasplit}\textstyle
F^A\Phi_A = (F_{A,cl}+\hat F_A)\Phi^A\ ,\qquad \hat F_A=\sum_{\vd>0}Q^\vd\qcord{\frac{\Phi_A}{1-qL}}_{\vd}\ ,
\ee
where $F_{A,cl}$ denotes the classical contribution\footnote{See eq.~\eqref{Fcl} below.} and $\hat F_A$ the quantum correlators. The precise form of the correlators depends on the choice of Chern--Simons terms. For zero effective levels $A_{ab}=0=B_a$ we find for the 1-point functions, to leading order in the degrees $Q_a$:
$$
\begin{aligned}
\hat F_0 &= -\tfrac{640
   (3 q-1) Q_1}{(q-1)^2}+\tfrac{4 (1-3 q) Q_2}{(q-1)^2} -\tfrac{16 \left(1761 q^4+5016 q^3+4122
   q^2-787\right) Q_1^2}{(q-1)^2 (q+1)^3}+\ldots \ , \\
\hat F_1 &= \tfrac{640 Q_1}{1-q}+\tfrac{32 \left(607 q^2+1254 q+667\right)
   Q_1^2}{(1-q) (q+1)^2}+\tfrac{640 Q_1 Q_2}{1-q}+\ldots \ ,\\
\hat F_2 &=\tfrac{4 Q_2}{1-q}
+\tfrac{640 Q_1
   Q_2}{1-q}-\tfrac{4 \left(q^2-2\right) Q_2^2}{(1-q) (q+1)^2}+\ldots \ .
\end{aligned}
$$
We will not give more explicit results at this point, since we found the universal formula eq.~\eqref{MCmm} below for the quantum correlators of $\IP^4_{1,1,2,2,2}[8]$, which holds also for all other Calabi--Yau threefolds that we studied so far (and with modifications also for Calabi--Yau manifolds of other dimensions, see sect.~\ref{sec:higherd}).

\subsection{Multi-Cover Formula for Calabi--Yau Threefolds \label{sec:mc3f}}
Below we propose a formula that gives the quantum correlators of the Calabi--Yau threefold $X$ in ordinary quantum K-theory at level zero in terms of the Gopakumar--Vafa invariants for $X$, or vice versa.  In the next section we derive the formula for the resolved conifold, which has only a single isolated curve of degree one. The general formula extrapolates the K-theoretic multi-cover formula of the conifold to higher degree maps, similarly to what has been done in the 2d context in ref.~\cite{AM93}.

We have explicitly checked the proposed multi-covering formula up to a certain degree in $Q$ for Calabi--Yau threefolds $X_3$ with up to three K\"ahler moduli, including the examples
$$
\begin{aligned}
&h^{1,1}=1: &&\IP_{1,1,1,1,1}^4[5]\,,\ \IP_{1,1,1,1,2}^4[6]\,,\ \IP^5_{1,1,1,1,1,1}[4,2]\,,\\
&h^{1,1}=2: &&\IP^4_{1,1,2,2,2}[8]\,,\ \IP^4_{1,1,2,2,6}[12]\,,\ \IP_{1,1,1}^2[3]\to X_3 \to \IP^2\,,\\
&h^{1,1}=3: &&\IP_{1,2,3}^2[6]\to X_3 \to \mathbb{F}_k\,,\ k=1,2\,.
\end{aligned}
$$
The last three examples are elliptic fibrations over the base $B=\IP^2$ and the Hirzebruch surfaces $B=\mathbb{F}_k$, respectively. In these cases we checked, that in the limit of large elliptic fiber one obtains the invariants for the local Calabi--Yau threefolds given by the cotangent bundles of the base $B$, e.g., $\cx O(-3)_{\IP^2}$ in the first case. The above examples, including the main example  $\IP^4_{1,1,2,2,2}[8]$, do not only contain isolated curves but also families. The numerical verifications give evidence for the proposed formula, but we do not have a mathematical proof. From the physics point of view, formula~\eqref{MCmm} says that the counting of BPS objects on the 3d world-volume and of 5d BPS objects in the target space differ only in the contribution of multi--covers in $n<3$-pt functions. It would be interesting to derive this difference from a membrane/target space duality.\\[-1.5ex]

\noindent {\bf Conjecture:}
The genus zero correlators at non-zero degree $\vd>0$ of ordinary quantum K-theory (at level zero) on Calabi--Yau threefold $X$ are related to the Gopakumar--Vafa invariants $\ngv_{\vd}$ for $X$ as
\be\label{MCmm}
\begin{aligned}
\hat F_0 &= p_2+\frac1{(1-q)^2}\big[(1-3q)\cx F+q\sum_at_a\cx F_{a}\big]_{t^{n>2}} \ , \\
\hat F_a &= p_{1,a}+\frac1{(1-q)}\big[\cx F_a\big]_{t^{n>1}}\ ,\quad  a=1,\hdots n_K\ ,\\
\hat F_A &=0\ ,\quad A>n_K\ .
\end{aligned}
\ee
Here 
\be\label{gw3p}
  \cx F(Q_ae^{t_a})= \sum_{\vd>\vec 0,n\geq 0} Q^\vd \, \frac{(\sum_j d_jt_j)^n}{n!}\, \sum_{r|\vd}\frac{\ngv_{\vd/r}}{r^3}\ ,\qquad
  \cx F_a =\p_{t_a}\cx F\ ,
\ee
is the  potential for the Gromov--Witten invariants, which depends only on the combinations $Q_ae^{t_a}$, and $[f]_{t^{n>n'}}$ denotes $f$ with the terms of degree $\leq n'$ in $t_a$ dropped. Moreover 
\be\label{pdefs}
\begin{aligned}
p_{1,a}=&\sum_{\vd>\vec 0} Q^\vd \, \sum_{r|\vd} \ngv_{\vd/r} \cdot 
  \Big\{ \tfrac{d_a}{r}\, (1-q) \tfrac{r(1-q^r)+q^r}{(1-q^r)^2} +  
   \tfrac{d_a}{r^2} (\sum_j t_j d_j)\,  \tfrac{1}{(1-q^r)}\Big\} \ , \\
p_{2}=&\sum_{\vd>\vec 0} Q^\vd \sum_{r|\vd} \ngv_{\vd/r} \cdot 
\Bigl\{
(1-q) \tfrac{r^2(1-q^r)^2-q^r(1+q^r)}{(1-q^r)^3}\\
&\hskip22ex+ \tfrac{\sum_j d_jt_j} r \frac{r(1-q^r)-q^r}{(1-q^r)^2}+\tfrac{(\sum_j d_jt_j)^2} {2r^2(1-q) }\Bigl\}\ .
\end{aligned}
\ee
The K-theoretic $n$-point functions with $n\geq 3$ are directly related to the Gromov--Witten prepotential $\cx F$ for $X$, as in the one modulus case considered in ref.~\cite{JM}. The coefficients of these $n$-point functions are, up to an overall power of $(1-q)$, the same as that of the cohomological expansion and integral.

On the contrary, the expansion of the $n<3$ point functions have non-integral coefficients in the cohomological theory, while they are integral in quantum K-theory. From the perspective of the 3d gauge theory, the integrality arises from the interpretation  of the supersymmetric partition function as a BPS index on the 3d world-volume. The integral 3d BPS invariants are encoded in the polynomials $p_{1,a}$ and $p_2$ of degree 1 and 2 in the deformations $t_a$, respectively. 

Eqs.~\eqref{MCmm},\eqref{pdefs} apply for the canonical choice $A=0=B$ for the Chern--Simons terms  in the 3d theory. For other Chern--Simons terms one obtains a similar relation between the two types of invariants, but the $q$-dependence of the multi--cover contributions  from higher degree maps is no longer the same as for degree one maps.
The classical terms in the $q$-periods are given in eq.~\eqref{Fcl}. It is straightforward to verify, that the 3d expressions reduce to the 2d period vector of $X$ in flat coordinates in the small radius limit.

\subsection{Local Conifold}
The simplest Calabi--Yau threefold with non-trivial quantum corrections is the local conifold described by the GLSM with charge vector $q_a=(1,1,\mm1,\mm1)$ and no hypersurface constraint. This is the non-compact threefold $X$ that contains a single rational curve of degree one with normal bundle $\cx O(-1)\oplus \cx O(-1)$. In order to extract correlators from the $I$-function we need to regularize the non-compact directions. This can be achieved by either introducing a real mass $m$ for the negatively charged fields of the GLSM, or by embedding the local geometry into a global one. We first discuss the mass deformation. The real mass $m$ becomes the weight with respect to the $U(1)$ flavor symmetry rotating the two chiral fields parametrizing the normal direction $\cx O(-1)\oplus \cx O(-1)$. On the level of geometry the $U(1)$ flavor symmetry becomes an $S^1$-action on the non-compact threefold $X$, which multiplies the fibers of the normal bundle $\cx O(-1)\oplus \cx O(-1)$ with a phase, and the fugacity $y=e^m$ of the flavor symmetry realizes the equivariant parameter of the equivariant K-group $K_{S^1}(X)$, which is given by
$$
  K_{S^1}(X) \simeq K(\mathbb{P}^1)[y,y^{-1}] \simeq \mathbb{Q}[P,y,y^{-1}]/(1-P)^2 \ .
$$
The strategy is now to compute equivariant correlator functions and then take the limit $y\to 1$. For the given geometry and in terms of the K-theoretic equivariant Euler class $e^K_{S^1} = (1 - y P^{-1})^2$ of the normal bundle $\cx O(-1)\oplus \cx O(-1)$ to $\mathbb{P}^1$ the equivariant holomorphic Euler characteristic becomes\footnote{For ease of notation we set $\beta=1$ in this subsection.}
$$
  \chi_{S^1}(X,E) = \chi\big(\mathbb{P}^1, \frac{E|_{\mathbb{P}^1}}{e^K_{S^1}} \big) \in \mathbb{Q}[y,y^{-1}] \ .
$$  
Taking the limit $y\to1$, which amounts to setting the real mass $m$ to zero, allows us to extract regularized holomorphic Euler characteristics of the non-compact threefold $X$. As a consequence the inner product is defined as 
\begin{equation} \label{eq:ConiMet}
  (\Phi_A,\Phi_B)_K = \lim_{y\to 1} \chi_{S^1}(X,e^K_{S^1}\Phi_A\Phi_B)=\chi(\mathbb{P}^1,\Phi_A\Phi_B)=\delta_{A+B,0}+\delta_{A+B,1} \ ,
\end{equation}
with $\Phi_0 = 1$ and $\Phi_1=1-P$.

Then the equivariant $I$-function of the symmetric theory for the conifold reads \cite{Giv15all}
\be\label{Icf}
\begin{aligned}
  I_{SQK}&=(1-q)\left( 1+\sum_{d=1}^{+\infty} Q^d \frac{\prod_{n=0}^{d-1}(1-y P^{-1}q^{-n})^2}{\prod_{n=1}^{d}(1- Pq^{n})^2}\right) \\
  &=(1-q) \left[ 1 + e^K_{S^1} \left( \sum_{d=1}^{+\infty}  \frac{Q^d}{q^{d(d-1)}P^{2d-2}(1-Pq^d)^2} + O(1-y) \right) \right] \ , \\
  &=(1-q) \Big[ 1 + e^K_{S^1} \big( I_0 \Phi_0 + I_1 \Phi_1 + O(1-y) \big) \Big] \ ,
\end{aligned}
\ee
where in the second and third line we have not displayed the terms of order $O(1-y)$, as these terms eventually vanish in the limit $y\to 1$, and with
$$
  I_0 = \sum_{d=1}^{+\infty}  \frac{Q^d}{(1-q^d)^2q^{d(d-1)}}\ ,\qquad 
  I_1 = \sum_{d=1}^{+\infty}  \frac{2Q^d}{(1-q^d)^2q^{d(d-1)}}\left(d - \frac1{1-q^d} \right) \ .
$$
Extracting the $\cx K_+$ part of these expressions we find for the input $t_0 \Phi_0 + t_1 \Phi_1$ of the symmetric quantum K-theory the expressions
$$
  t_0 = (1-q) \,e^K_{S^1} \, \sum_{n=1}^{d-1} n\,q^{d(n-d)} \ , \quad
  t_1 = (1-q) \,e^K_{S^1} \, \sum_{n=1}^{d-1} n(2d-n-1)\, q^{d(n-d)} \ ,
$$  
such that the $I$-function becomes
\begin{multline} \label{eq:Isp}
  I_{SQK} = (1-q)+ (t_0 \Phi_0 + t_1 \Phi_1) \\
   + (1-q) e^K_{S^1} \sum_{d=1}^{+\infty} Q^d \Big[ \Phi_0 \tfrac{d(1-q^d)+q^d}{(1-q^d)^2} \\
   + \Phi_1 \tfrac{d(d-1)(1-q^d)^2-2 q^d}{(1-q^d)^3} 
   + O(1-y) \Big] \ ,
\end{multline}
where the terms in the square bracket reside in the $\cx K_-$ part.

We observe that the input $t_0 \Phi_0 + t_1 \Phi_1$ is proportional to equivariant K-theoretic Euler class $e^K_{S^1} \sim O(1-y)$. Hence, the input of the $I$-function vanishes in the limits $y\to1$, and therefore only contributes to the equivariant terms $O(1-y)$ in $\cx K_-$ in eq.~\eqref{eq:Isp} (which we have not spelled out explicitly). Upon removing this input with a suitable transformation~\eqref{gtfs}, we therefore arrive at the $I$-function
\begin{multline*} 
  I = (1-q)+ (1-q) e^K_{S^1}\sum_{d=1}^{+\infty} Q^d \left[ \Phi_0 \tfrac{d(1-q^d)+q^d}{(1-q^d)^2}\right.\\
   \left.+ \Phi_1 \tfrac{d(d-1)(1-q^d)^2-2 q^d}{(1-q^d)^3}+ O(1-y) \right] \ .
\end{multline*}
From this expression together with the metric~\eqref{eq:ConiMet} we readily read off in the limit $y\to 1$ the 1-pt functions\footnote{Note that, in principal, the permutation symmetric input~$t_0\Phi_0+t_1\Phi_1$ yields higher point degree zero correlators, which --- due to the $Q$-dependence of this input --- potentially contributes to $\cx K_-$ at the same order in $Q$ as the extracted 1-pt correlators. However, in the limit $y\to 0$ all these contributions vanish.}  
\begin{equation}\label{cfcc1}
\begin{aligned}
  \qcor{\frac{\Phi_1}{1-qL}}_d&= (1-q)\frac{d(1-q^d)+q^d}{(1-q^d)^2}  \ , \\
  \qcor{\frac{\Phi_0}{1-qL}}_d&= (1-q)\frac{d^2(1-q^d)^2-q^d(1+q^d)}{(1-q^d)^3}  \ .
\end{aligned}
\end{equation}
We can now generate the non-trivial input $t \Phi_1$ for the ordinary quantum K-theory by acting with the transformation $\exp\left({\frac{t (1 - P q^\theta)}{1-q}}\right)$ according to ref.~\cite{GivER}, which yields the $I$-function
$$
  I_{QK}(t) = (1-q) +  K_0(Q,q,t) \Phi_0 +  K_1(Q,q,t) \Phi_1+ e^K_{S^1}\, O(1-y)  \ ,
$$
where $K_\ell(Q,q,t)$, $\ell=1,2$, decompose into the input $K_\ell^+(Q,q,t)$ and the correlator contributions $K_\ell^-(Q,q,t)$. A straightforward but tedious computation yields the (for us relevant) input
$$
  K_0^+(Q,q,t) = e^K_{S^1} \sum_{d=2}^{\infty}\sum_{n=1}^{+\infty} \frac{t^nQ^d}{n!} k^+_{d,n}(q) \ , \quad
  K_1^+(Q,q,t) = t + e^K_{S^1} ( \ldots ) \ ,
$$
with
$$
  k^+_{d,n}(q) = 
  \begin{cases} 
    (d-1)+O(1-q) & \text{for } d\ge2, n=1 \ , \\
    \frac{d^{n-2}(d-1)(2d-n+2)}2+O(1-q) & \text{for }d\ge2,n>1 \ , \\
    0 & \text{else} \ .
   \end{cases}
$$
Furthermore, the correlator contributions read
$$
\begin{aligned}
  K_0^-(Q,q,t) &=e^K_{S^1} \sum_{d=0}^{+\infty} Q^d \left[ (1-q)\tfrac{d(1-q^d)+q^d}{(1-q^d)^2} + \tfrac{t}{1-q^d} 
    + \sum_{n=2}^{+\infty} \tfrac{d^{n-2}t^n}{n!(1-q)} \right] \ ,\\
  K_1^-(Q,q,t) &=e^K_{S^1} \sum_{d=0}^{+\infty} Q^d \Bigg[ (1-q)\tfrac{d(d-1)(1-q^d)^2-2 q^d}{(1-q^d)^3} + \tfrac{t(d(1-q^d)-1)}{(1-q^d)^2} 
     + \tfrac{t^2(d-1)}{1-q}   \\
     &\qquad  + \sum_{n=3}^{+\infty} \frac{t^n}{n!} \left(\tfrac{d^{n-3} (n-2)}{(1-q)^2} +\tfrac{d^{n-3}(n-3) (n-2)}{2 (1-q)}
        -\tfrac{d^{n-2} (2-n-2 d n+n^2)}{2 (1-q)}\right) \Bigg]  \ . \end{aligned}
$$
Due to the $Q$-dependent terms in the input in $K_0^+$, we see that the $n$-pt function in $K_0^-$ at degree $d$ combines with $2$-pt functions at degree $0$, such that we obtain with the metric~\eqref{eq:ConiMet} the equations
$$
\begin{aligned}
   \sum_{n=0}^{+\infty} \frac{t^nQ^d}{n!} \qcor{\frac{\Phi_1}{1-qL},\Phi_1^n}_d &= \left. K_0^-(Q,q,t) \right|_{Q^d}  \ , \\
   \sum_{n=0}^{+\infty} \frac{t^nQ^d}{n!} \qcor{\frac{\Phi_0}{1-qL},\Phi_1^n}_d & = \left.K_0^-(Q,q,t)\right|_{Q^d}  +\left.K_1^-(Q,q,t)\right|_{Q^d}  \\
     &-\sum_{n=2}^{+\infty} \frac{t^n Q^d}{(n-1)!} k_{d,n-1}^+(1) \qcor{\frac{\Phi_0}{1-qL}, e^K_{S^1}\Phi_0,\Phi_1}_0 \ .
\end{aligned}
$$
The appearing degree zero correlator reduces to the ordinary quantum K-theory correlator of the point, such that we get with eq.~\eqref{eq:ConiMet}\footnote{As a result of the degree zero correlator only $k_{d,n-1}^+(q)$ at $q=1$ contributes to $\cx K^-$. Furthermore, we note that the correlator $\qcor{\frac{\Phi_0}{1-qL}, e^K_{S^1}\Phi_1,\Phi_1}_0$ vanishes.}
$$
  \qcor{\frac{\Phi_0}{1-qL}, e^K_{S^1}\Phi_0,\Phi_1}_0 = \chi_{S^1}(X,e^K_{S^1}\Phi_1) \qcor{\frac{1}{1-qL}, 1,1}_\text{pt} = \frac{1}{1-q} \ .
$$  
These two identities allow us to derive the $(n+1)$-pt functions of the ordinary quantum K-theory, which turn out to be
\begin{equation} \label{cfcc2}
\begin{aligned}
 &\qcor{\frac{\Phi_1}{1-qL},\Phi_1}_d=\tfrac{1}{1-q^d}\ ,
   &&\ \qcor{\frac{\Phi_1}{1-qL},\Phi_1^n}_d= \tfrac{d^{n-2}}{1-q}\ \text{for }n>1\ ,\\
 &\qcor{\frac{\Phi_0}{1-qL},\Phi_1}_d=\tfrac{d(1-q^d)-q^d}{(1-q^d)^2}\ ,
   &&\ \qcor{\frac{\Phi_0}{1-qL},\Phi_1^2}_d=\ \tfrac{1}{1-q}\ ,\\
 &\qcor{\frac{\Phi_0}{1-qL},\Phi_1^n}_d=d^{n-3}\tfrac{1-q(3-n)}{(1-q)^2} &&\!\!\!\text{for }n>2 \ .
\end{aligned}
\end{equation} 
The correlators in eqs.~\eqref{cfcc1} and \eqref{cfcc2} sum up to the expression~\eqref{MCmm} for $n_K=1$ and $\ngv_{k>0}=0$.

A geometric regularization of the non-compact directions, which is  in the spirit of local mirror symmetry, is to embed the local conifold geometry into a global Calabi--Yau threefold and study its decompactification limit. A simple compactification of the local conifold is the elliptic fibration $X$ over the Hirzebruch surface $\mathbb{F}_1$ described by a 3d GLSM with gauge group $U(1)^3$ and matter fields with charge vectors $$
l_1=(\mm 6;3,2,1,0,0,0,0),\  l_2=(0;0,0,\mm2,1,1,0,0),\ l_3=(0;0,0,\mm1,0,\mm1,1,1).
$$
In the limit of large K\"ahler classes $Q_{1,2}\to0$, the elliptic fiber and the $\mathbb{P}^1$ fiber of $\mathbb{F}_1$ decompactify and the local geometry of the compact $\IP^1$ is the one studied above. Consider the $q$-hypergeometric series $\cx I$ which solves the system of difference equations eq.\eqref{DiffEqgen} for the compact manifold $X$. The leading term in the decompactification limit is $\cx I_{lim}=\lim_{Q_{1,2}\to 0}\cx I = Q_1^{\eps_1}Q_2^{\eps_2}I_\textrm{local}(Q_3)$. The only non-trivial difference equation in this limit arises from the difference operator $\DL_3=(1-p_3)^2-Q_3(1-p_3^{-1}p_2)(1-p_3^{-1}p_1p_2^{-2})$. Applying $\DL_3$ to $\cx I_{lim}$ and dividing by $Q_1^{\eps_1}Q_2^{\eps_2}$ we obtain the difference equation
$$
\DL_\textrm{lim}\ I_\textrm{local}(Q)=0, \qquad \DL_\textrm{lim}= (1-p)^2-Q(1-p^{-1}y_1)(1-p^{-1}y_2)\ .
$$
Here  we used $Q=Q_3$, $p=p_3$, $y_1=q^{\eps_2}$ and $y_2=q^{\eps_1-2\eps_2}$.
The difference operator $\DL_\textrm{lim}$ is the same as the difference operator for the equivariant theory of the local conifold for real masses $y_1,y_2$ of the non-compact directions. In particular, spezializing to $y=y_1=y_2$, $\DL_\textrm{lim}$ annihilates the equivariant $I$-function~\eqref{Icf}.\footnote{To compensate for the fact that eq.\eqref{Icf} is written without the overall factor $P^{-\ln Q/\ln q}$, $p$ has to be taken to be $p=Pq^\theta$.} This shows the equivalence of the two regularizations. 

One can also compute directly in the non-equivariant limit. Noticing that the non-equivariant limit of the relation $e_{S^1}^K(1-P)^2=0$ is $(1-P)^4=0$, one arrives at a degree four difference operator for the non-equivariant conifold of the form
$$
\DL_\textrm{lcf} = (1-p)^2(1-zp^{-2})(1-p)^2\ .
$$
The solutions of $\DL_\textrm{lcf}$ are the coefficients of the $q$-hypergeometric series $\cx I$ in eq.\eqref{Icf} at $y=1$ in an expansion up to order $(1-P)^3$. From here, the computation in the non-equivariant theory proceeds as in the previous examples and leads to the same result eqs.\eqref{cfcc1},\eqref{cfcc2}.

The conifold example captures the contribution for a rigid rational curve of degree one with normal bundle $\cx O(-1)\oplus \cx O(-1)$. The general formula eq.~\eqref{MCmm} predicts, that the K-theoretic multi--cover contributions from other curves are of the same universal form,  up to extra combinatorical factors of the degree $d_a$ (which have to be consistent with the 2d limit). E.g. the $\IP^4_{1,1,2,2,2}[8]$ contains a family of rational curves of degree $\vd=(0,1)$ with normal bundle $\cx O(-2)\oplus \cx O(0)$. The local model for this curve can be obtained as the non-compact limit $Q_1\to 0$ of the compact threefold, and it gives the same multi--cover formula as for the rigid curve.

\subsection{K-Theoretic Ring}
The quantum correlators \eqref{MCmm} determine the quantum deformation of the multiplication rings in quantum K-theory.
Adding the classical terms to the quantum correlators, the $q$-period vector in the basis $\{\Phi_0,\Phi_{a},\Phi^a,\Phi^0\}$ is
\be\label{qpv}
\Pi (Q,t) = \begin{pmatrix}
\qq\\
\hat t_a\\
F_{a,cl}+p_{1,a}+\frac1{\qq}\big[\cx F_a\big]_{t^{n>1}}\\
F_{0,cl}+p_2+\frac1{\qq^2}\big[(1-3q)\cx F+q\sum_at_a\cx F_{a}\big]_{t^{n>2}}
\end{pmatrix}.
\ee
Here $\qq = 1-q$ and 
\bea\label{Fcl}
F_{a,cl}&=& \kappa_{abc}\left(\frac1{2\qq}\that_b\that_c-\qq \frac{L_b}{2}\delta_{bc}\right)+\sum_{A>n_K} t_A\chi_{Aa}\ ,\nonumber\\
F_{0,cl}&=&\kappa_{abc}\left(\frac1{3!\qq^2}\that_a\that_b\that_c-\frac{\delta_{ac}}2L_a\that_b
-\qq\frac{\delta_{ab}\delta_{ac}}3L_a\right)\\
&&+\frac{\chi_{AB}}{2\qq} \that_A\that_B-\qq \frac{L_a}2\chi_{aa}+\sum_{A>n_K} t_A\chi_{A0} \ ,\hskip-2cm\nonumber
\eea
where the small and capital indices run over the sets $a=1,\ldots,n_K$ and $A=0,\ldots,\dim(K(X))-1$, respectively, and the shifted variables $\hat t_A$ are given by
$$\textstyle
\hat t_A = t_A-\delta_{Aa}\qq L_a\, ,\qquad L_a = \frac{\ln Q_a}{\ln q}\ .
$$
The period matrix $\pi$ defined from $\Pi $ is
\be
\pi = \begin{pmatrix}\  \pi_0\\[1mm]\pi_{1,a}\\[1mm]\pi_{2,\mu}\\[1mm]\ \ \pi_*\end{pmatrix}=
\begin{pmatrix}\hskip18pt \Pi^T\\[1mm] \qq\p_{t_a}\Pi^T\\[1mm]\qq\p_{t_\mu}\Pi^T\\[1mm]\qq\p_{t_*}\Pi^T\end{pmatrix}\ ,\nonumber
\ee
for the deformation $t = \sum_{A>0}t_A\Phi_A = t_a\Phi_a+t_\mu \Phi_{n_K+\mu}+t_*\Phi_{1+2n_K}$. Here we use $t_a$, $a=1,\ldots n_K$, $t_\mu$, $\mu=1,\ldots n_K$ and $t_*$ to denote the deformations in the directions of the basis elements with minimal cohomological degree two, four and six, respectively.

The flatness of the connection in the $t$-directions can be expressed in terms of the linear differential equations for the period matrix
\be
\qq \p_{t_A}\pi(Q,t) = C_A(Q,t)\pi(Q,t)\ .\nonumber
\ee
Here $C_A(Q,t)$ are the matrices of structure constants. Starting from \eqref{qpv} one finds 
\be
\qq \p_{t_a}\begin{pmatrix}\pi_0\\\pi_{1,b}\\\pi_{2,\mu}\\\pi_*\end{pmatrix} 
= \begin{pmatrix}
0&\delta_{ac}&0&0\\
0&0&\hat C_{ab}^\nu&\hat c_{ab}\\
0&0&0&\hat\chi_{a\mu}\chi_{0*}^{-1}\\
0&0&0&0
\end{pmatrix}\ \begin{pmatrix}\pi_0\\\pi_{1,c}\\\pi_{2,\nu}\\\pi_*\end{pmatrix} \ ,
\ee
and 
\be
\qq \p_{t_\mu}\pi = \begin{pmatrix}
0&0&\delta_{\mu\nu}&0\\
0&0&0&\hat\chi_{b \mu}\chi_{0*}^{-1}\\
0&0&0&0\\
0&0&0&0
\end{pmatrix}\  \pi\ ,\qquad 
\qq \p_{t_*}\pi = \begin{pmatrix}
0&0&0&1\\
0&0&0&0\\
0&0&0&0\\
0&0&0&0
\end{pmatrix}\ \pi \ ,\nonumber
\ee
where 
\bea
\hat C_{a b }^\nu \hat\chi_{\nu c} &=& \kappa_{abc}+\cx F_{abc}=: C^{GW}_{abc}(Qe^t)\ ,\nonumber\\
\hat c_{ab}\chi_{0*}&=& [\cx F_{ab}]_{t^{n>0}}-\cx F_{abc}t_c+\chi_{ab}-\hat C_{ab}^\nu \chi_{\nu 0}+\qq p_{2,ab}\ .\nonumber
\eea
Here $C^{GW}_{abc}$ are the structure constant of the GW theory, which depend only on the combinations $Q_ae^{t_a}$ of the parameters $(Q,t)$. The $n_K\times n_K$ matrix $(\hat\chi)_{c\nu}:= (\Phi_c,\Phi_{\nu+n_k})_K$ is invertible  and $\chi_{0*}=(\Phi_0,\Phi_{1+2n_k})_K$ is non-zero, i.e. the above relations can be solved for $\hat C$ and $\hat c$. 

From the above it follows, that the only 3d product with non-trivial quantum deformation is 
\def\qp{\otimes}
$$
\Phi_a\qp \Phi_b = \hat C_{ab}^\nu \Phi_\nu + \hat c_{ab}\Phi_*\ ,
$$
where $\hat C_{ab}^\nu=\hat\chi^{\nu c}C^{GW}_{abc} $ are determined by the structure constants of the cohomological  theory, and the second term $\hat c_{ab}$ depends in addition on the $n\leq 2$-point correlators expressed in terms of the cohomological invariants  in eq.~\eqref{MCmm}.

The matrices $C_A$ satisfy the flatness relations $[C_A,C_B]=0=\p_{t_A}C_B-\p_{t_B}C_A$, which follow from the WDVV equations of quantum K-theory \cite{GivWDVV,IMT}, or, from the point of the underlying 3d field theory,  the 3d $tt^*$ equations \cite{CV13}.

\section{Other Dimensions}  \label{sec:higherd}
Dimensions other than three are also interesting for several reasons. For dimension less than three, i.e. for $T^2$  and K3 manifolds, we find that the $I$ function in ordinary quantum K-theory computed as above is the classical one
$$\textstyle
I(t_A) = (1-q)+t_A\Phi_A+\sum_{A>n_K}F_{cl}^A(t)\Phi_A\ , 
$$
where the last term is zero for $T^2$. There is still interesting non-perturbative information in the symmetrized, or more generally permutation equivariant, theory. In particular, for all dimensions, the 3d vortex sum $\cx I(Q,q)$ is non-trivial and computes the coefficients of the ordinary 2d mirror map in terms of the integral degeneracies of 3d BPS states, as discussed in sect.~8.3 of ref.~\cite{JM}.

For dimension higher than three, the cohomological Gromov--Witten invariants can be still be computed from the entries $\Pi_\gamma^{\textrm{ln}^2}(Qe^t)$ of the 2d period vector\footnote{I.e., the 2d version of the solution vector eq.~\eqref{Piq}.} with double logarithmic behavior \cite{GMP}. The index $\gamma$ runs over a basis of $H^4(X,\mathbb{Z})\cap H^{2,2}(X)$. After normalization, the potentials $\cx F_\gamma\sim \Pi_\gamma^{\textrm{ln}^2}$ have an expansion for large K\"ahler moduli of the form \cite{PM4f, KLRY}
\be\label{defn4f}
\cx F_\gamma(Qe^t)= \cx F_{\gamma,cl}(Qe^t)+\sum_{\vd>\vec 0,n\geq 0} Q^\vd \, \frac{(\sum_j d_jt_j)^n}{n!}\, \sum_{r|\vd}\frac{\ngv_{\gamma,\vd/r}}{r^2}\, ,
\nonumber\ee
where the classical contribution $\cx F_{\gamma,cl}(Qe^t)$ is a degree two polynomial in $\ln(Qe^t)$. The invariants $\ngv_{\gamma,\vd}$ defined by this expansion are integral invariants associated with a 4-cycle $C_\gamma\in H_4(X,\mathbb Z)$. The $h_{2,2}(X)$ potentials $\cx F_\gamma$  replace the Gromov--Witten potential \eqref{gw3p} of the threefold case. They are related to the quantum corrected 3-point correlators for the operators $\phi_\alpha,\phi_\beta\in H^{1,1}(X,\mathbb Z)$  and $\phi_{\gamma}\in H^{d-2,d-2}(X,\mathbb Z)$ as 
\be
C_{\alpha\beta\gamma}=\phi_\alpha\wedge \phi_\beta \wedge \phi_\gamma = \p_{t_\alpha,t_\beta}\cx F_\gamma\ .\nonumber
\ee

Computing correlators of the ordinary quantum K-theory for various Calabi--Yau $d$-folds, we find a very similar structure for the multi-cover contributions as for the threefold case in sect.~\ref{sec:mc3f}. The $I$ function for a deformation $t=t_a\Phi_a$ has again the general form in eq.~\eqref{I3f}. Moreover, let  $\Phi_\gamma=K_\gamma+\hdots $ be an element of the basis \eqref{Pdef} with $K_\gamma\in H^{2,2}(X,\mathbb{Q})$ and
$$
\hat F_\gamma=\sum_{\vd>0}Q^\vd\qcord{\frac{\Phi_\gamma}{1-qL}}_{\vd}\ ,\qquad \gamma = n_K+1,\hdots,n_k+h^{2,2}(X)\ ,
$$
the quantum correlators associated with these elements. Then all our computations are consistent with the following conjectural expression of the correlators in terms of the invariants $\ngv_{\gamma, \vd}$ defined in eq.~\eqref{defn4f}:
\bea\label{MCmm4f}
\hat F_\gamma &=& p_{1,\gamma}+\frac1{(1-q)}\big[\cx F_\gamma\big]_{t^{n>1}}\ ,\  a=1,\hdots n_K\ ,\\
p_{1,\gamma}&=&\sum_{\vd>\vec 0} Q^\vd \, \sum_{r|\vd} \ngv_{\gamma,\vd/r} \cdot \Bigl\lbrace
 (1-q) \frac{r(1-q^r)+q^r}{(1-q^r)^2}\ + \ 
\frac{1}{r} (\sum_j t_j d_j)\,  \frac{1}{(1-q^r)}\Bigr\rbrace \nonumber 
\eea
This is essentially the same formula as for the correlators $\hat F_a$ in the threefold case, with the replacement $d_a\ngv^\textrm{d=3}_\vd \to \ngv^\textrm{d$>$3}_{a,\vd}$.

As an example we consider the non-compact toric Calabi--Yau four-fold $X_4$ corresponding to the charge vectors \def\td{{\tilde d}}
$$
l_1 = (0;-3,1,1,1,0,0),\qquad l_2 = (0;1,0,0,-1,1,-1)\ .
$$
The generalized $q$-hypergeometric series  $\cx I$ for $X_4$, for zero effective Chern--Simons terms and with $\td_a = d_a-\eps_a$, is
$$
\cx I = \sum_{d_1,d_2\geq 0}z_1^{\td_1}z_2^{\td_2} \,
\tfrac{\Gamma_q(1-\eps_1)^2\Gamma_q(1-\eps_2)\Gamma_q(1+\eps_2)\Gamma_q(1+3\eps_1-\eps_2)\Gamma_q(1-\eps_1+\eps_2)}
{\Gamma_q(1+\td_1)^2\Gamma_q(1+\td_2)\Gamma_q(1-\td_2)\Gamma_q(1-3\td_1+\td_2)\Gamma_q(1+\td_1-\td_2)}\ .
$$
The correlators in ordinary quantum K-theory are obtained from the vortex sum $\cx I$ following the steps $\cx I \to I_{SQK}(t) \to I(0) \to I_{QK}(t)$ as outlined in sect.~\ref{sec:MCSec} for the threefold case. The two independent $q$-periods $F_\gamma$ can be chosen to have classical pieces $F_{3,cl}=\alpha_3 t_1^2$ and $F_{4,cl}= \alpha_4 (t_1+3t_2)^2$ with constants $\alpha_{3,4}$; in this basis $F_3(t_1,Q_1)$ does not depend on $t_2$ and $Q_2$ and agrees with the $q$-period of the Calabi--Yau threefold $X_3$ for $\alpha_3 = -\frac 1 6$. The 1-pt correlators obtained from $F_4$ with $\alpha_4=-\frac 1 {12}$ are listed in Table~\ref{tab:CorX4}.
\begin{table}[h]
\begin{small}\begin{center}$$
\vbox{\offinterlineskip
\halign{\vrule\strut~~$#$~~&\vrule ~~$#$~~&~~$#$~~&~~$#$~~&~~$#$~~\vrule\cr
\noalign{\hrule}
\phantom{Q\over Q}&Q_2^1&Q_2^2&Q_2^3&Q_2^4\cr
\noalign{\hrule}
Q_1^0&-1 & \frac{q^2-2}{(q+1)^2} & \frac{2 q^3-3}{\left(q^2+q+1\right)^2} & \frac{3 q^4-4}{(q+1)^2
   \left(q^2+1\right)^2} \cr
Q_1^1&2 & 1 & 1 & 1 \cr
Q_1^2&-5 & -\frac{2 q (3 q+4)}{(q+1)^2} & -3 & -\frac{5 q^2+8 q+2}{(q+1)^2} \cr
Q_1^3&32 & 21 & \frac{2 \left(9 q^4+16 q^3+27 q^2+18 q+12\right)}{\left(q^2+q+1\right)^2} & 20 \cr
Q_1^4&-286 & -\frac{5 \left(35 q^2+72 q+38\right)}{(q+1)^2} & -153 & 
*
\cr
\noalign{\hrule}}}
$$\end{center}\end{small}\vskip-0.5cm
\caption{1-pt correlators for $X_4$ that compute disk invariants  in the $q\to 1$ limit.} \label{tab:CorX4}
\end{table}

In the context of the 2d/GW correspondence, it is known \cite{PMdual} that the genus zero GW invariants of $X_4$ compute Ooguri--Vafa {\it disk} invariants for a certain family of Lagrangian branes $L$  in the threefold $X_3=\cx O(-3)_{\IP^2}$ described in refs.~\cite{OV,AV0,AKV}, with $t_1$ and $t_2$ measuring the size of a sphere and a disk in the three-dimensional geometry $(X_3,L)$, respectively. Since the GW theory of $X_4$ can be obtained by sending the $S^1$ radius in the 3d theory to zero, the $q\to1$ limit of the 3d QK  invariants in Table~1 also reproduces the 2d disk invariants for the geometry $(X_3,L)$. It is natural to ask about an interpretation of the QK invariants in terms the geometry $(X_3,L)$, which would lift the 2d open/closed duality between $(X_3,L)$  and $X_4$ to the 3d theory. Such an interpretation should involve quantum K-theory on the moduli space of Riemann surfaces with boundary on the mathematical side and it would be interesting to study this further, perhaps along the lines of ref.~\cite{KL}, where a mathematical definition for the cohomological disk invariants has been given.\\[0cm]

\noindent{\bf Acknowledgments:} 
We would like to thank Bumsig Kim, Urmi Ninad and Yongbin Ruan for discussions.
The work of P.M. is supported by the German Excellence Cluster Universe.\\[-1cm]

\bigskip

\providecommand{\href}[2]{#2}\begingroup\raggedright\endgroup
\end{document}